\documentclass[12pt]{article}
\pdfoutput=1
\usepackage{putex}
\usepackage{feyn}
\usepackage[vcentermath]{youngtab}
\usepackage{subfig}
\usepackage{lscape}

\usepackage{graphicx}
\usepackage{epstopdf}
\usepackage{enumerate}
\usepackage{cite}
\usepackage{tensor}
\usepackage{slashed}
\usepackage{amsmath}
\usepackage{amssymb}
\usepackage{mathrsfs}
\usepackage{lgrind}

\usepackage{amsmath,amssymb,braket}

\usepackage{bbm}

\usepackage{hyperref}

\numberwithin{equation}{section}

\def\dd{{4}}

\newcommand {\be} {\begin {equation}}
\newcommand {\ee} {\end {equation}}

\newcommand {\bes} {\begin {equation*}}
\newcommand {\ees} {\end {equation*}}

\newcommand{\eps}{\epsilon}

\def\CH{{\cal H}}

\newcommand{\beq}{\begin{equation}}
\newcommand{\eeq}{\end{equation}}

\def\be{ \begin{equation} }
\def\ee{ \end{equation} }

\def \be {\beta}

\def \beq { \begin{equation}}
\def \eeq {\end{equation}}

\renewcommand\Re{\operatorname{Re}}
\renewcommand\Im{\operatorname{Im}}

\usepackage{braket}
\begin{document}

\preprint{ITEP-TH-}

\institution{MIPT}{Institutskii per, 9, Moscow Institute of Physics and Technology, 141700, Dolgoprudny,
Russia}
\institution{ITEP}{B. Cheremushkinskaya, 25, Institute for Theoretical and Experimental Physics,
117218, Moscow, Russia}
\institution{UC}{Universit\`a degli Studi dell'Insubria - Dipartimento DiSAT, Via Valleggio 11 - 22100 Como - Italy}
\institution{INFN}{INFN, Sez di Milano, Via Celoria 16, 20146, Milano - Italy}
\institution{PU}{Department of Physics, Princeton University, Princeton, NJ 08544}

\title{Ultraviolet phenomena in AdS self-interacting quantum field theory}

\authors{Emil~T.~Akhmedov\worksat{\MIPT,\ITEP}, Ugo Moschella\worksat{\UC, \INFN} and Fedor K.~Popov\worksat{\PU, \MIPT,\ITEP}}

\abstract{{We study the one-loop corrections to the four--point function in the Anti de Sitter space--time for a $\phi^4$ field theory.  
Our calculation shows the existence of non--local counterterms which however respect the AdS isometry. Our arguments are quite general and applicable to other AdS field theories. We also explain why calculations in Euclidean and Lorentzian signatures should differ even at the leading order in non globaly hyperbolic manifolds.}}

\date{}
\maketitle
 
 \section{Introduction}

Quantum field theories in spaces with constant curvature have received considerable attention during the last four decades. The interest in spaces with negative curvature arouse twenty years ago because of a deep connection between gravitational theories in their "bulks" and conformal field theories on their boundaries \cite{Maldacena:1997re,Gubser:1998bc,Witten:1998qj}. The interest in spacetimes having positive curvature is older as they play a major role in contemporary cosmology \cite{Guth:1980zm,Linde:1981mu,Albrecht:1982wi}. 
According to the standard cosmological model, the Universe is believed to have passed through a quasi-de Sitter phase at the early stages of its evolution and to approach again a de Sitter geometry today in its {\em dark age}. The study  of quantum fields on the de Sitter background may also shed some light on the cosmological constant problem \cite{Krotov:2010ma,Polyakov:2012uc,Polyakov:2009nq,Akhmedov:2013xka,Akhmedov:2012pa,Akhmedov:2013vka}. 

Constant curvature spaces have maximal isometry groups. Their highly symmetric status allows for the explicit calculation of propagators and loop corrections, a fact which is not at all obvious in general gravitational backgrounds. 
In de Sitter (dS) space there is a peculiar infrared (IR) behavior of the loop corrections to the correlation functions even for massive fields \cite{Krotov:2010ma,Polyakov:2012uc,Polyakov:2009nq,Akhmedov:2013xka,Akhmedov:2012pa,Akhmedov:2013vka}. 
On the other hand, the ultraviolet (UV) behavior of quantum fields in spaces of constant curvature is believed to be the same as in flat space, at least at the leading order of the semiclassical expansion, following the idea that the high energy modes should not change even in the presence of  a large quasiclassical gravitational field. 

In this paper we calculate the one-loop renormalization of the four--point function in various constant curvature spaces, both in Lorentzian and  Euclidean signature. We show that in Anti de Sitter space the UV renormalization is quite different from what one does in flat space and for a very simple reason. 

Two basic assumptions are used in the standard UV renormalization procedures in flat space: first, the UV modes, having a very short wavelength, should not be sensitive to the boundary conditions; in other words a UV renormalization procedure should be local. Second, calculations can be done in the Euclidean signature and then analytically continued to the Minkowski spacetime. 

The singular support of the propagators in the Euclidean manifold is a set of isolated points. The UV renormalization obviously depends on such singularities which are in turn closely related to the properties of the UV modes. On the other hand, in the Minkowski manifold the singular support of propagators is made of light--cones emanating from source points. If all the propagators have only one singular point (source) this fact causes no trouble. 

Problems may, however, arise when propagators have more than one source (once more, in Lorentzian signature). Even when the source does not belong to the space--time, its light--cone can penetrate into it (see \cite{Akhmedov:2017hbj} for an early observation of this phenomenon). This is the situation encountered in flat space in presence of a perfect mirror, a mirror which reflects all momenta equally well. Similar in spirit, though with some important differences, situation appears in global Minkwoskian Anti de Sitter space.  

The structure of the paper is as follows.
In the second section to set up the notations we provide the $x$--space derivation of the beta function. Then, to clarify our main point on a simple example, we exhibit analogous calculations in flat space in  presence of an ideal mirror.
In the third section we summarize some properties of the two--point correlation functions in spaces of constant curvature. 
In the fourth section we calculate the four--point function renormalization in Anti de Sitter (AdS) space. In the Appendix we explain the physical origin of the analytical properties of the two--point functions in AdS space.

Loops in Lobachevsky space (Euclidean AdS) for scalar and higher spin fields were considered in  \cite{Sleight:2017cax}, \cite{Sleight:2017pcz}. Even earlier, in \cite{Akhmedov:2012hk} loops in AdS space were calculated with the emphasis on their IR properties.

\section{UV renormalization in presence of a mirror}

\begin{figure}
    \centering
    \includegraphics[scale=0.8]{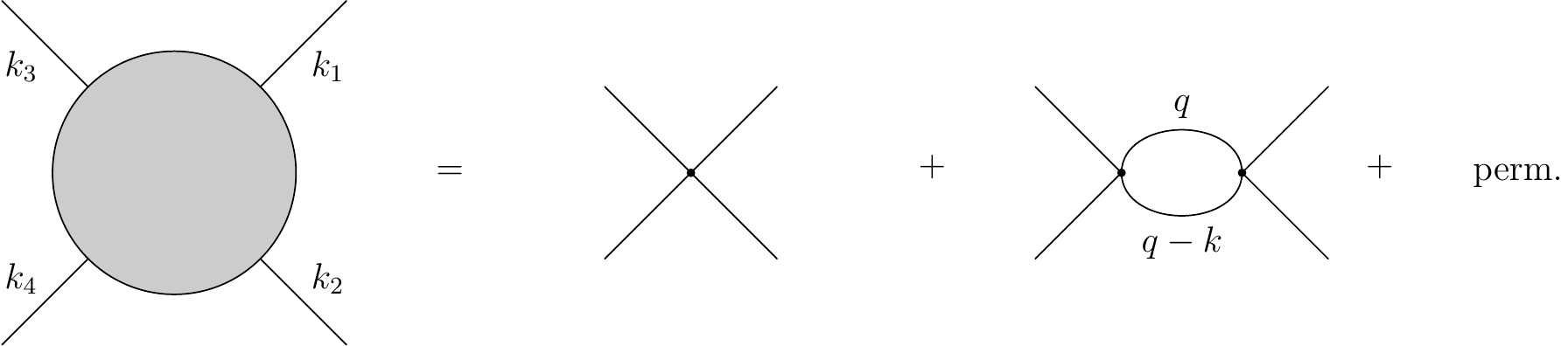}
    \caption{The one-loop corrections to the four-point correlation function in the interacting scalar field theory.}
    \label{fig:scat}
\end{figure}

In our conventions the signature of the metric is $(+,-,-,-)$. 
We consider a scalar field theory with quartic selfinteraction:

\beq \label{scalar4theory}
S = \int d^4 x \, \sqrt{-g} \, \left[\frac12\left(\partial_\mu \phi\right)^2 - \frac12 m^2 \phi^2 - \frac{\lambda}{4!} \phi^4 \right].
\eeq
The one loop contribution to the effective action of the theory --- see the diagram in Fig. \ref{fig:scat} --- is  given by the following expression:
\begin{gather}
\label{deltaS}
\Gamma^{(4)} = - i \, \frac{3\lambda^2}{2 (4\pi^2)^2} \, \int d^4x \int d^4 y \, \phi^2(x) \phi^2(y) \, F^2\left(x-y\right),  
\end{gather}
where 
\beq\label{flatfey}
F(x) \approx \frac{1}{4\pi^2} \frac{i}{x^2-i\eps},
\eeq
is the most singular part of the Feynman propagator in position space  when $ x^2 \to 0$.

We may extract the  leading divergent contribution in (\ref{deltaS})  by changing the  variables 
$ x^\mu = X^\mu + \frac{z^\mu}{2}$, $y^\mu = X^\mu - \frac{z^\mu}{2}$, $\mu = 0,1,2,3$ and by diagonally expanding $\phi^2\left(X + z/2\right) \, \phi^2\left(X - z/2\right)$: 
\beq
\Gamma^{(4)}= - i \, \frac{3\lambda^2}{2 \, (4\pi^2)^2} \, \int d^4X  \, \phi^4(X) \, \int d^4z \frac{1}{\left(z^2 - i \eps\right)^2} + {\rm UV \, \, finite \,\, terms} \label{singA}
\eeq
The $z$--integral in the last expression provides the standard logarithmic UV divergence and can be handled  by a UV cutoff $1/M$ (or any other type of regularization). 

To illustrate our main point on a simple example we continue by considering the theory (\ref{scalar4theory}) in flat space-time but in the presence of an ideal mirror placed at $x_3=0$. The ideal mirror reflects all the modes equally well,   irrespectively of their momenta. This is expressed  by the  boundary condition $\left.\phi\right|_{x_3 = 0} = 0$ at the mirror. 

However a physical mirror is definitely transparent to very high energy modes.  On general physical grounds one can expect that, if $a$ is a characteristic interatomic distance of the material of the mirror, a mode whose wavelength $k$ is much larger than ${1}/{a}$ will not see the mirror at all. Such a situation can be modeled by a potential barrier which reflects some of the modes and is transparent to the other ones \cite{Akhmedov:2017hbj}. Let us consider for instance the field equation
\beq 
\left[\Box + m^2 \right] \, \phi = \alpha \, \delta(x_3) \, \phi.
\eeq
It is not difficult to see that the modes with $k\alpha \ll 1$ are reflected while those with $k\alpha \gg 1$  pass through.  In more realistic situations mirrors will have several windows in momentum space of transparancy and reflection.

The most singular part of the Feynman propagator in  presence of an ideal mirror  is the following distribution
\beq\label{mirrprop}
F(x,y) \approx \frac{i}{4\pi^2} \frac{1}{s - i\eps} - \frac{i}{4\pi^2} \frac{1}{\bar{s} - i\eps}, \quad \text{where} \quad s=(x-y)^2\quad \text{and}\quad \bar{s}=(x-\bar{y})^2,
\eeq
and $\bar{y}$ is the mirror image of the source point $y$. Note that $y=\bar{y}$ if and only if $y$ is on the mirror surface.  
The propagator \eqref{mirrprop} can be obtained by the method of image charges (see e.g. \cite{landau1960course}), where one uses the reflection  $\phi(\bar{x}) = -\phi(x)$ for fields and currents. 

By looking at the expression (\ref{mirrprop}) one can also immediately grasp the difference between the Euclidean and Lorentzian cases. In fact, in Euclidean signature, $s$ vanishes only when $x=y$ and $\bar{s}$ only when $x = \bar{y}$. But the point $\bar{y}$ does not belong to the portion of space-time  that we are considering. The mirror limits the region of integration in (\ref{deltaS}) to the points such that $x_3>0$ and $y_3 > 0$.
Hence, inside the loops in Euclidean signature $\bar{y}$ plays no role. On the other hand, in Lorentzian signature, $s$ and $\bar{s}$ vanish on the light--cones whose tips are $y$ and, respectively, $\bar{y}$. Therefore, even though $\bar{y}$ does not belong to the space-time manifold its light--cone penetrates into it (see fig. \ref{fig:mirror}). That leads to a different UV renormalization at the leading order, as we will see below.

The situation we find here is similar to the one encountered in AdS quantum field theory. Indeed, the  AdS manifold is also not globally hyperbolic and the AdS  Feynman propagators have two singularities too (see below). However, there are some important differences. First,  the second singularity arises from  points that do belong to the global AdS manifold or its covering. Second, as we will recall below, (Wightman's) propagators in AdS space are analytic functions of the hyperbolic distance $\zeta$ while the propagator \eqref{mirrprop} is not associated to an analytic function of the interval $s$ alone but it also depends on $\bar{s}$.

\begin{figure}
    \centering
    \includegraphics{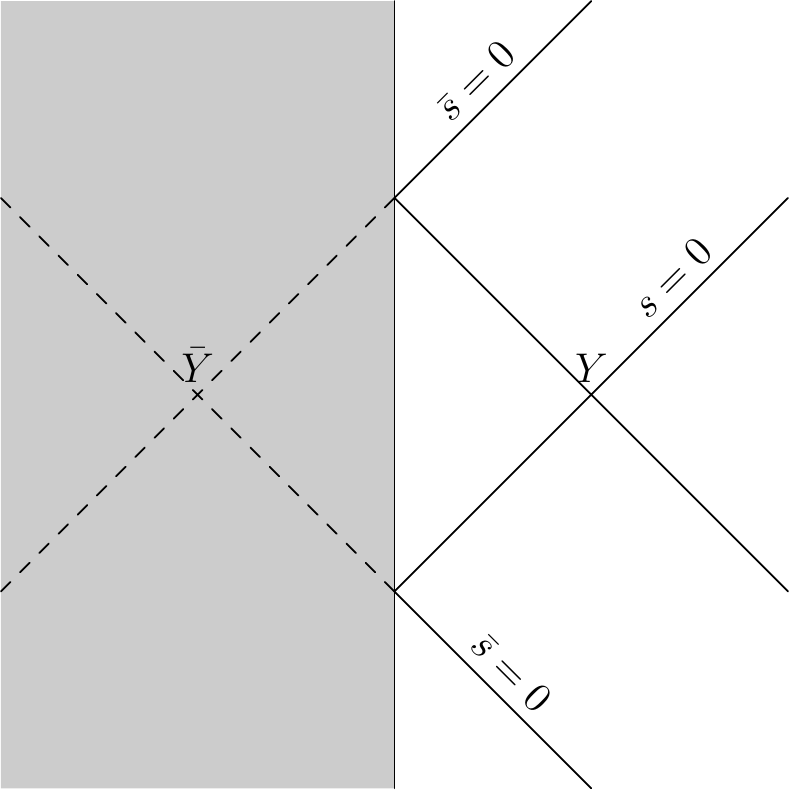}
    \caption{Light cones in the case of the ideal mirror.}
    \label{fig:mirror}
\end{figure}

Moreover, to obtain Lorentzian expressions starting from the Euclidean manifold one has to perform analytical continuations either in the Mandelstam variables in momentum space or in the geodesic distance in position space. But when  a mirror is present, the propagator is not a function of such variables alone. This is true both in position and in  momentum space, as it results from the above decomposition  (\ref{mirrprop}) of the propagator. It should be however said that the existence of the second singularity regards only an ideal mirror -- one reflecting even very high momenta. A real physical mirror would be transparent to high momenta and the second singularity in (\ref{mirrprop}) would simply not be there.

Let us now calculate the contributions to the effective action (\ref{deltaS}) arising from the diagram in Fig. \ref{fig:scat}. The first term in (\ref{mirrprop}) provides the same contribution as in Eq. (\ref{singA}). As regards the second term we have to do the following diagonal expansion: $ x^\mu - \bar{y}^\mu = z^\mu$, $x^\mu + \bar{y}^\mu = 2\, X^\mu$, $\mu = 0,1,2,3$. Then, the effective action contains the following term:
\beq\label{4.7}
\Gamma_{\bar{s} = 0}^{(4)} = - i \, \frac{3\lambda^2}{2 \, (4\pi^2)^2} \, \int d^4X  \,\int_{z_3 \geq \left|X_3\right|} d^4z \frac{\phi^2(X+z/2) \, \phi^2(\bar{X}-\bar{z}/2)}{\left(z^2 - i \eps\right)^2}.
\eeq
The status of this divergence is now different: even though  $z^2$ may vanish the components of four--vector $z$ are generically not small (see fig. \ref{fig:mirror}) and the diagonal expansion of $\phi^2(X+z/2) \, \phi^2(\bar{X}-\bar{z}/2)$ cannot be performed. Consequently, this  divergence cannot be subtracted by a local counterterm. Both the singularities of $F$ at $s=0$ and $\bar{s} = 0$ do contribute to the UV divergence of the integral on the right hand side, while the mixed terms contribute finite expressions.

Such non--local UV divergent contributions actually appear for other choices of the boundary conditions at the mirror, including either the Neumann or the  mixed conditions. We postpone the discussion of the consequences of such a situation to the concluding section.

It is worth stressing here the following point. The Feynman prescription in Eq. (\ref{singA})  makes the spacetime integral divergent, as usual. However, if in that equation one replaces the Feynman propagator with the Wightman function  then the integral is obviously vanishing:

\beq\label{closedcont}
\int d^4z \frac{1}{\left[\left(z_0 - i \epsilon\right)^2 - \vec{z}^2\right]^2} = 0. 
\eeq
If for any reason the second singularity is of Wightman's type as opposed to the $i \epsilon$ prescription used in (\ref{mirrprop}) then the second UV divergence is not there. This arguments of course can be applied to the eq. (\ref{4.7}) only if $\phi^2(X+z/2) \, \phi^2(\bar{X}-\bar{z}/2)$ is a holomorphic function of $z_0$ in the lower complex half-plane.

However, it should be stressed that here we consider a non globaly hyperbolic manifold and the issue of the boundary conditions is highly important in such a case. The   perfect mirror boundary condition immediately rules out any prescription different from (\ref{mirrprop}). Other types of boundary conditions may lead to conclude that the second singularity does not contribute.

\section{Propagators in spaces of constant curvature}

Let us consider the linear space ${\mathbb R}^5$   endowed with the Euclidean scalar product 
\beq \label{scalarprod}
 X \cdot X'=  \sum_{i=0}^4 X_i X'_i .
 \eeq 
The 4--dimensional sphere of radius $R$ embedded in ${\mathbb R}^5$ is the quadric 
\beq \label{sphereq}
S_4= \left \{ X\in {\Bbb R}^5, \ \ X \cdot X=  X^2 = R^2 \right\}.
\eeq 
The geodesic distance  $d\left(X,X'\right)$ between two points on the sphere is related to their scalar product (\ref{scalarprod}) in the ambient space as follows:
\beq 
 X\cdot X'= {R^2} \cos \frac{ d (X,X') }{R} .\label{geospher}
 \eeq
$d\left(X,X'\right)$ is the length of the arc of  the equator  obtained by intersecting the sphere with the plane containing $X$, $X'$ and the origin of the ambient space.

The real sphere is a real submanifold of the complex sphere
\beq \label{sphereqc}
CS_4=\left \{ Z\in {\Bbb C}^5, \ \ Z \cdot Z= R^2 \right \},
\eeq 
embedded into the complex linear space ${\Bbb C}^5$ endowed with the scalar product
$
Z \cdot Z'=  \sum_{i=0}^4 Z_i Z'_i .
$
Since $R^2$ is a real number,   writing  $Z_j= X_j+i Y_j$ allows to split Eq. (\ref{sphereqc})  into the following two conditions:
\beq \label{sphereq2r}
CS_4=\left \{ X+iY \in {\Bbb C}^5,\ \ X^2-Y^2 = R^2, \ \ X\cdot Y = 0\right\}.  
\eeq 
The real sphere (\ref{sphereq}) is not the only remarkable (real) submanifold of $CS_4$. If we restrict the first coordinate to be purely imaginary,  $Z_0 = i Y_0$ and keep the remaining four coordinates real we obtain the 4--dimensional dS spacetime, identified to  the one--sheeted real hyperboloid 
\beq \label{ds}
dS_4= \{-Y _0^2 + X_1^2 + X_2^2 + X_3^2 +X_4^2 = R^2\}
\eeq 
which has Lorentzian signature; in this vein, the real sphere  (\ref{sphereq}) can be thought as the Euclidean dS manifold $S_4=EdS_4$.  

On the contrary, we may keep real the first coordinate  $Z_0 = X_0$ and take the remaining four coordinates $ Z_j = i Y_j, \  j=1,\ldots, 4$ purely imaginary; 
this gives  the 4--dimensional Euclidean AdS space (identical to the Lobachevsky space) 
\beq \label{eads}
EAdS_4= \{X _0^2 - Y_1^2 - Y_2^2 - Y_3^2 - Y_4^2 = R^2\},
\eeq 
a manifold which has Euclidean signature; it is one of the sheets (say $X_0 \geq R$) of the two--sheeted real hyperboloid. We could have arrived at same equation by the replacement  $R\to iR$ in Eq. (\ref{ds}). 

Finally we may keep real the first and the last coordinates $Z_0 = X_0$, $Z_4 = X_4$,  take the remaining three coordinates  purely imaginary,   $ Z_j = i Y_j, \  j=1,2, 3$,  
and get the 4--dimensional (Lorentzian) AdS space:  

\beq \label{ads}
AdS_4= \{X _0^2 - Y_1^2 - Y_2^2 - Y_3^2 + X_4^2 = R^2\}.
\eeq 
Correspondingly, the isometry groups of the real spaces under consideration are $SO(5,{\Bbb R})$,  $SO(1,4,{\Bbb R})$ and $SO(2,3,{\Bbb R})$; they are all real subgroups of the complex group $SO(5,{\Bbb C})$.

The invariant scalar product 
\begin{equation}
\zeta = \frac{Z\cdot Z'}{R^2}
\label{invsc}
\end{equation} 
 takes real values when restricted to any of the above manifolds. 
In Lobachevsky space this is also called the hyperbolic distance and is again related to the geodesic distance as follows: 
\begin{equation} { Z\cdot Z'}= {R^2} \cosh \frac{ d (Z,Z') }{R} \geq R^2, \label{geohyp}
\end{equation}
(for  $Z_0Z'_0>0 $ so that  $\zeta \geq 1$). Here $d\left(Z, Z'\right)$ is the length of the arc of the hyperbola obtained by intersecting $EAdS_4$ with the  plane containing $Z$, $Z'$  and the origin of the ambient space.

In the Lorentzian cases, however, not always the  two points in Eq. (\ref{invsc}) can be joined by an arc of geodesic.  
For instance in the dS case, where $Z= (i Y_0, \vec X)$, the geodesic distance between two points is implicitly defined as in Eq. (\ref{geohyp})  
for $\zeta>1$  (i.e for $Z$ and $Z'$ timelike separated), as in Eq. (\ref{geospher})  
{for} $|\zeta|\leq 1$
and not defined at all for $\zeta< -1$. As before $d\left(Z, Z'\right)$ is the arc length of either the hyperbola or the ellipse joining the two points $Z$ and $Z'$. Similar remarks apply to  the AdS case. Both in dS and AdS $\zeta$ takes all the values between plus and minus infinity. 

Consider now a two-point function $W(Z,Z') = w(\zeta)$ depending only on the invariant variable $\zeta$ 
and defined in some domain of the complex $d$-dimensional sphere.
We may extend the function $w$ to the ambient space by homogeneity of degree zero 
by considering the function $w_{ext}\left(\frac{ Z\cdot Z'} {\sqrt {Z^2}{\sqrt{Z'}^2}}\right)$, where $Z$ and $Z'$ are no more constrained to the complex sphere.  
The complex Laplace operator applied to $w_{ext}$ (in general dimension $d$) coincides with the the Laplace-Beltrami operator on the sphere applied to $w$: 
\begin{equation}
\left. 
\nabla^2 w_{ext} \right|_{CS_d} = \left. \frac  \partial {\partial Z_\mu}\frac {\partial w_{ext}}  {\partial Z^\mu} \right |_{CS_d} = 
\left(1-\zeta^2 \right)\, \partial_\zeta^2 w(\zeta)- d \, \zeta \, \partial_\zeta w(\zeta).
\end{equation}
The   following equation in the invariant variable:
\beq\label{wightprop}
\left[\left(1-\zeta^2 \right)\, \partial_\zeta^2 - d \, \zeta \, \partial_\zeta +\sigma(d-1-\sigma) \right] \, w(\zeta) = 0 
\eeq
represents therefore either the Laplace or the Klein-Gordon equation (possibly also with a delta source  at the RHS).

The parameter $\sigma$ is related to the mass as follows: 
\beq \pm \left(m \, R\right)^2 =\sigma(d-1-\sigma) \eeq
i.e.
$$\sigma =\frac{d-1} 2 +  \ \sqrt{\left(\frac{d-1}{2}\right)^2 \pm  \left(m \, R\right)^2}= \frac{d-1} 2 +  \nu. $$
The plus sign is chosen for the AdS and Lobachevsky cases, while the minus sign is chosen for the dS case and the sphere. However in both cases $\left(m\, R\right)^2$ can also take negative values, as we recall in Appendix \ref{apb}.

Changing  to either $ \frac{1+\zeta}{2}$ or $ \frac{1-\zeta}{2}$ gives to Eq. (\ref{wightprop}) the standard hypergeometric form. Correspondingly the general solution may be written in general dimension $d$  as follows:

\begin{equation}
w(\zeta) = \frac{ {2}^{\frac{d-2}2} A_\pm \, {}_2F_1 \left(\frac{1} 2 +  \nu, \, \frac{1} 2  -  \nu; \frac{d}{2}; \frac{1+\zeta}{2}\right) }{\left({1-\zeta}{}\right)^{\frac{d-2}2} }+  \frac { 2^{\frac{d-2}2} B_\pm {}_2F_1 \left(\frac{1} 2 +  \nu, \, \frac{1} 2 -  \nu ; \frac{d}{2}; \frac{1-\zeta}{2}\right)}{\left({1+\zeta}\right)^{\frac{d-2}2} } \label{propgen}
\end{equation}
where ${}_2F_1$ is the hypergeometric function and we used Kummer's relation. 

$A_\pm$  and $B_\pm$ are some complex coefficients that are chosen according with the sign of  $\Im \zeta$. 
There are indeed cuts on the real axis of the complex $\zeta$-plane which come from the quantum commutation relations; one has to specify suitable analyticity properties for $w$ which depend on the chosen geometry (i.e. dS or AdS) in such a way that the upper and lower boundary values of  the Wightman  functions in (\ref{propgen})  coincide at spacelike separated pairs.  
In particular, in the dS case two (real) points $Z$ and $Z'$ are time-like separated when $(Z-Z')^2<0$ and therefore in the complex  plane  of the invariant variable $\zeta$ there is cut (at least)  on the positive reals starting from $\zeta=1$. 
On the other hand, in the AdS case two (real) points  $Z$ and $Z'$ are time-like separated when $(Z-Z')^2>0$ and therefore the cut is opposite to the previous i.e. it is the half-line $(-\infty ,1]$. In some special periodic cases, when $\nu = \sqrt{\left(\frac{d-1}{2}\right)^2 +  \left(m \, R\right)^2}= \frac 12 , \frac 32 , \frac 52 \ldots $, the two contributions in (\ref{propgen}) compensate and cut reduces to the interval  $[ -1,1]$.

The inverse image -- in the complex sphere at fixed $Z'$ --  of the first singularity $\zeta=1$ is the intersection of the complex sphere $Z^2=1$ with the complex cone $(Z-Z')^2= 0$. This surface includes coinciding points on either the sphere or the Lobachevsky space  and lightlike separated pairs on either the dS or the AdS manifold. On the other hand the singularities of the second term in (\ref{propgen}) are on  the intersection of the complex sphere $Z^2=1$  with the complex cone $(Z+Z')^2= 0$; the latter is  the cone having its tip at $-{Z}'$, the antipodal point  of $Z'$. 

Every point $Z$ on the complex sphere has an antipode $-Z$. This reflection gives the antipodal points also on the real submanifolds we are considering in this paper (on the Lobachevsky space the antipodal point belongs to the other sheet of the two--sheeted hyperboloid).  
The crucial difference is the following:  while in the dS case antipodal pairs are spacelike separated in the AdS case they are not. Actually, in the AdS case all the timelike geodesics issued from an event $Z$ focus  at the antipodal point $-Z$ ; this fact remains true also on a general covering of the (real) AdS space. This is the reason why in the AdS case the second singularity in Eq. (\ref{propgen}) is always present.

In the dS case there is a special choice of vacuum, namely the Bunch--Davies or Euclidean vacuum \cite{Bunch:1978yq} which is maximally analytic \cite{bros,bros2,bros3} and corresponds to the choice $A_+=A_-$, $B_\pm=0$ in (\ref{propgen}). 
The maximal analyticity properties implies that the Schwinger function on the sphere is the analytic continuation of the two-point function on the real dS manifold and that there are no singularities in between. 
However there are other choices, namely the so called alpha-vacua {\cite{Spindel,Mottola:1984ar,Allen:1985ux}} (see \cite{Akhmedov:2013vka} for a review) which are dS invariant (at least at tree--level) but  they are not analytic precisely because of the presence of the additional singularity. 

In the AdS case the situation is a little more involved due to the presence of a non trivial topology of the real manifold. As we show in the Appendix, the Feynman propagator  can be represented both in global AdS and in its covering $\widetilde{AdS}$ manifold as follows:

\begin{eqnarray}  \label{FAdS}
&& F_\nu\left(X,X'\right) = \cr && -\frac{i \, \Gamma\left(\frac {d-1}2-\nu \right) \Gamma\left(\frac {d-1} 2 +\nu\right)}{ 2\,{(2\pi)^{\frac{d}2}} \Gamma(\frac {d}2)}
\left[  \left(\frac{1}{\xi  -1 +i\epsilon}\right)^{\frac {d-2}2}  \phantom{|}_1F_2\left(\frac 1 2-\nu ,\frac 1 2+\nu,\frac {d}2;\frac {1+ \xi+ i \epsilon} 2 \right)\right. \cr && - \left. 
e^{- i(\nu-\frac 1 2)\pi} 
\left(\frac{1}{\xi+1+i\epsilon}\right)^{\frac {d-2}2}  \phantom{|}_1F_2\left(\frac 1 2-\nu,\frac 1 2+\nu,\frac {d}2;\frac {1-\xi - i\epsilon}2  \right)\right]  \label{prop0}
\end{eqnarray}
Here $X,X'$ denote two events of the real AdS manifold (\ref{ads}) and $\xi$ denotes their  invariant scalar product (\ref{invsc})  (see also (\ref{ambientmetricads})).
Both contributions on the RHS of this expression separately grow when $\xi \to \infty$, i.e. as the distance between $X$ and $X'$ is increasing. But in the linear combination appearing here the growing contributions compensate each other and the  resulting expression is decaying with the distance. That agrees with \cite{Avis:1977yn}, \cite{kent2013quantum} (see also \cite{Polyakov:2007mm} and \cite{Akhmedov:2009ta}).  The corresponding Wightman function is maximally analytic \cite{bbgm,bem}. Note also that

\begin{equation}
(X-X')^2-i \epsilon = 2R^2(1 - (\xi+i\epsilon))
\end{equation}
and therefore the prescription $\xi \to \xi + i \epsilon$ corresponds to the standard prescription in Minkowski space. 
Details on the analiticity properties of AdS propagators are given in Appendix. The relation of the above   $i \eps$ prescription with the local time ordering of the AdS manifold and the global time ordering of its covering is discussed below and in Appendix.

\section{Counterterms in AdS field theories}
\begin{figure} 
    \centering
    \includegraphics[scale=1]{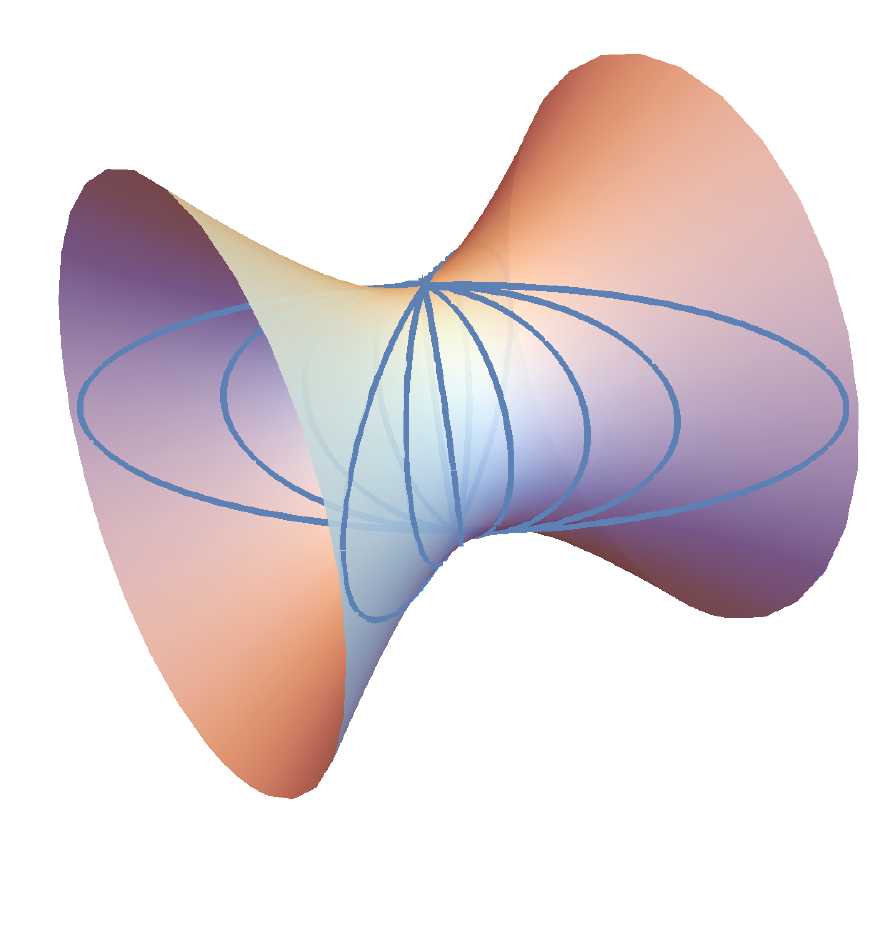}
    \caption{The timelike geodesics issued from a point $Z$ focus at the antipodal point $-Z$. The boundary acts somehow like a (thick) mirror for massive particles. }
    \label{fig:adsgeo}
\end{figure}
The AdS manifold is non globally hyperbolic for two reasons: it has closed timelike curves and has a boundary at spacelike infinity. The first problem may be "cured" moving to the universal covering  which however remains non globally hyperbolic because it still has a boundary at infinity.  Whatever is the case, the geodesics issued from any point of the manifold (or of its covering) focus after half a period at the antipodal event (see fig. \ref{fig:adsgeo}). At the quantum level, this fact is related to the presence of the second singularity in the Wightman function (\ref{propgen}).  When $\xi^2\to 1$ in four dimensions,  Eq. (\ref{FAdS}) provides the following leading singularities of the AdS Feynman propagator\footnote{In the Euclidean AdS (Lobachevsky) manifold $\xi \geq 1$ and the second singularity cannot be attained (recall that $-Z'$ the point antipodal to $Z'$ belongs to the opposite  sheet). Moreover, because the Lobachevsky space has Euclidean signature the singularity is only attained at coincident points $Z=Z'$ and we do not have to care about the contributions to the effective action arising from the second singularity (\ref{FAdS}).}:
\beq\label{adsproptaylor}
F(\xi) \approx -\frac{i}{8\pi^2 \, \left(\xi-1+i\eps\right)} - 
\frac{e^{-i\pi \nu}}{8\pi^2 \, \left(\xi + 1 + i\eps\right)}.
\eeq
The same $i\eps$ prescription applies in the fundamental sheet both at $\xi \to -1$ and at $\xi \to 1$ (see Appendix \ref{apb} - we have taken the radius of AdS to be one $R=1$).
This fact is a consequence of the the spectral condition and of the local time ordering of the AdS manifold.
The chosen $i \epsilon$ prescription implies that the propagator solves the following inhomogeneous Klein--Gordon equation with an extra delta source on the RHS: 
\beq
\left[\Box + m^2\right] F(X,X') = 4 \, \pi \, \delta(X,X') + 4 \, \pi \, i \, e^{-i \, \pi\, \nu} \, \delta(X, -X').\label{fundeq}
\eeq
 We can now  calculate the diagram in fig. \ref{fig:scat}. Here is the form of the loop correction for the simplest case  $\nu = 1/2$ in the global AdS manifold:
\begin{eqnarray}\label{4.3}
\Gamma^{(4)} \propto \lambda^2 \, \int d^5X\, \delta\left(X^2 - 1\right) \,
\int d^5X'\, \delta\left({X'}^2 - 1\right) \, \phi^2\left(X\right) \, \phi^2\left(X'\right) \times \nonumber \\ \times \left[\frac{1}{\left(X-X'\right)^2 - i\eps} - 
\frac{1}{\left(X + X'\right)^2 - i\eps}\right]^2.
\end{eqnarray}
The square of the first pole can be treated in the standard way  and leads to the same renormalization as in flat spacetime. The square of the second pole is different and leads to  divergences of a new type. Finally, the cross terms lead to less singular contributions.
\begin{figure}[h]
    \centering
    \includegraphics[scale=1.2]{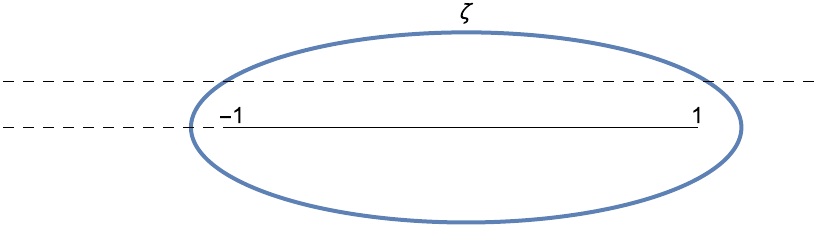}
    \caption{The cut plane of the complex $\zeta$ variable is the maximal domain of analyticity of two-point functions satisfying the spectral condition. In the periodic cases the extra dotted cut disappears. The dotted line $\Im \zeta = \epsilon$ is the projection on the complex $\zeta$ plane of the domain of integration in Eq. (\ref{4.3}).}
\label{cutplane}
\end{figure}

Let us examine now the general case of a field with bare mass parameter $\nu$. Its Wightman two-point function, considered as a function of the invariant variable $\zeta$, is analytic in the  covering $\widetilde\Theta$ of the
cut-plane $\Theta = \{{\mathbb C} \setminus[-1,1]\}$, see Fig. \ref{cutplane}. 
The singularities are located at $\zeta = \pm 1$. 
Fixing a point somewhere on the real covering manifold $\widetilde {AdS}$ and moving the other in (a suitable domain of) the complex manifold $\widetilde {CAdS}$, every time that $\zeta$ accomplishes a round trip around the two singularities the two point function  gains a phase $w(\zeta) \to e^{2\pi i (\nu-\frac 12 )} \, w(\zeta)$.

In the periodic case, the domain of integration in the loops projects on the dotted line in fig. \ref{cutplane}.  
We may choose to keep the same contour which lies entirely in the fundamental domain also in the non periodic case. Alternatively, we may exploit the global time ordering of the covering manifold and construct the chronological propagator accordingly; the result given is  in Eq. (\ref{cpcov}), coincides with the propagator (\ref{prop0}) in the fundamental sheet. The propagator (\ref{cpcov}) solves the  Klein-Gordon equation (\ref{fundeq}) on the covering manifold without the supplementary delta--function on the RHS.
However, it is unclear how to use that propagator in loop integrals over the covering manifold $\widetilde{AdS}$. Its quasi periodic structure seems to lead to an infinite result even after UV regularization, and extracting a meaningful (non ad hoc) finite result in such a situation is problematic.

Thus, if we insist in integrating over the fundamental domain of $\widetilde{AdS}$ also in non periodic cases, i.e. for generic values of $\nu$, as in Eq. (\ref{4.3}) we have to  introduce a new counter--term in the Lagrangian \eqref{scalar4theory}:

\beq\label{newterm}
\Delta \mathcal{L} = \frac{\gamma \, e^{- \, 2\, \pi\,i \, \nu}}{4} \phi^2(Z) \phi^2(-Z),
\eeq
with a complex coefficient depending on the   mass parameter  and a new coupling constant $\gamma$.
 
Here we have described the situation in either the global AdS manifold or its covering. In the Poincar\'e patch the situation demands a separate treatment. First, one has to modify Eq. (\ref{4.3}) by restricting the regions of integration to a chosen patch. Of course such a choice violates the AdS symmetry. However, for the $i\epsilon$ prescription as in eq. (\ref{adsproptaylor}) the isometry is preserved in the loops,  while is violated for any other choice \cite{Akhmedov:2013vka}. Then, for the (\ref{adsproptaylor}) choice the situation is similar to the case of the perfect mirror: even though the antipodal point  $-X$ does not belong to the same patch containing a generic point $X$, the light--cone of $-X$ obviously  penetrates into the patch.

It should be noticed that, although the term (\ref{newterm}) is non-local, it nevertheless respects the AdS symmetry.
A somewhat similar situation has been  described in \cite{Banks:2002nv} for the dS alpha--vacua.  There is however an important difference: the loop corrections to the alpha-vacua do not respect the dS isometry due to infrared contributions \cite{Polyakov:2012uc}, \cite{Akhmedov:2013vka}. This means once more that in the dS case the tree--level ground state is the Bunch--Davies vacuum, which does not share such problems.
In conclusion, the AdS UV behavior is quite different from the Minkowskian one and should be considered carefully.

\section{Conclusion}

The main result of this paper is that the UV renormalization of the four--point function in  AdS space--time generates non--local counter terms that respect the isometry group. 

This phenomenon may be regarded as another revelation of the so called UV/IR mixing, similar to the one found in the case of electric field and global dS space \cite{Polyakov:2012uc}.  
It shows a crucial difference between quantum fields in flat, dS and AdS spaces. In dS space we have the standard UV behavior for the properly chosen basis of modes, while the IR effects are strong and quite different from those in flat space (see \cite{Akhmedov:2013vka} for a review). On the other hand, in the AdS space (in Lorentzian signature)  no new problem arises in the IR regime \cite{Akhmedov:2012hk} but UV physics is quite different from flat space. 
As in the case of a perfect mirror we encounter a UV divergence which cannot be removed by a local counterterm. 

The  question now is whether these effects are physical or it is possible to reformulate quantum field theory in such a way that they do not show up. For instance one may try to use compactly supported field algebras.  
It is worthwhile to remark that, unlike the case of the perfect mirror, the situation in global AdS is yet under control. There seems to be no local measurement which allows to detect terms such as (\ref{newterm}) and their presence does not destroy the renormalizability of the theory (although they do affect the beta--function). Obviously one can discard the situation of the perfect mirror as unphysical, but still accept the effects discussed in this paper for AdS space as physical (despite the presence of the complex coupling constant).

Another possibility would be to define  quantum field theory in AdS space via the analytical continuation from  Lobachevsky space, i.e. from the Euclidean manifold. However, such an analytical continuation does not allow to address the issues of e.g. non--stationary phenomena within the AdS/CFT correspondence (at least beyond the $1/N$ approximation). Indeed, the analytical continuation from the Euclidean signature does not provide correct answers in \emph{non--stationary situations} already in flat space \cite{landau2013statistical}. In non--stationary situations correlation functions are not functions of the hyperbolic distance anymore (see e.g. \cite{Akhmedov:2013vka} for a similar discussion in dS).
 
\section*{Acknowledgments}

We are grateful to I.R. Klebanov, J.Maldacena, A.M.Polyakov, M.Taronna and S.Theisen for valuable discussions. ETA would like to thank the quantum gravity group of AEI, Golm for the hospitality during the final stage of the work on this project. ETA is partially supported by the Russian state grant Goszadanie 3.9904.2017/8.9 and RFBR 16-02-01021. FKP is partially supported by the grant RFBR 18-01-00460 А. 

\appendix

\section{Analyticity properties of the two--point functions in the complex AdS space}
\label{apb}

\subsection{Geometry}

Let us introduce the following scalar product in $\mathbb R^{d+1}$
\begin{equation}
\xi(X,X') = X\cdot X'= {X_0} X'_0+ {X_\dd}X'_\dd - X_1 X'_1 - {X_2} X'_2 -{X_3} X'_3 -\ldots .
\label{ambientmetricads}
\end{equation}
(in this section the "dot" notation that  was previously used for the Euclidean scalar product indicates the product (\ref{ambientmetricads})).  The $d$-dimensional AdS spacetime can be identified with the quadric
\begin{equation}
AdS_{d} = \{ X \in \mathbb R^{d+1}, \ \ X^2= {  X} \cdot{X}=R^2\}.
\end{equation}
The real AdS interval is then obtained by restricting the ambient space interval

$$ds^2 = d{X}_0^2 +d{X}_\dd^2  -d{X}_1^2 -d{X}_2^2 -d{X}_3^2 -\ldots . $$ to AdS itself. The AdS isometry group is the pseudo-orthogonal group $G =SO(2,d-1 ,{\Bbb R})$. The complex AdS manifold $$CAdS_{d} = \{ Z = X+iY \in \mathbb C^{d+1}, \;\;Z^{2}= R^2\} $$ obviously coincides with the complex sphere $CS_4$. 
 $Z = X+iY$ belongs to $CAdS_{\dd}$ if and only if $$X^2 - Y^2 = R^2  \ \ {\rm and}   \ \ X\cdot Y = 0.$$ 
In the following we will put  $R=1$.
The  complex AdS manifold is parametrized by giving complex values 
to the  coordinates below:
\begin{equation} Z = Z(t, \psi,{\omega})=
\left\{\begin{tabular}{lclcll}
$Z^{0}= X^0+iY^0 $  &=& $\cosh \psi \sin t $ & \cr
$Z^{i} = X^{i} + iY^{i} $  &=& $\sinh \psi \  {\omega}^i $                   & ${ i=1,2,\ldots,d-1}$ & \cr
$Z^{d} = X^{d} +iY^{d}$ &=&$\cosh \psi \cos t $&
\label{sphericcoordinates}
\end{tabular}\right.
\end{equation}
with ${\omega}^2 = 1.$ 
The real AdS manifold is parametrized by restricting the  coordinates to real values.
This coordinate system allows to easily describe also the universal covering spaces $\widetilde {AdS}$  and
 $\widetilde {CAdS}$ by unfolding the periodic  time coordinate $\Re{t}$.
The covering $\widetilde G$ of the group $G=SO(2,d-1)$ is introduced in a similar way.
We will use the symbols $X$, $Z$, etc., also to denote points of the coverings.

Two events of the real manifold $AdS$ are space-like separated when
\begin{equation}(X- X')^2 = 2 - 2 X\cdot  X' <0.\end{equation}
Since AdS and $\widetilde {AdS}$ are transitively generated by the action
of $G$ and $\widetilde G$ on the  base point
$$B=(0,\ldots,0,1)$$ we  may specify the notion of space-like separation in the covering space
$\widetilde {AdS}$ as follows: let  $X,X'\in \widetilde{AdS}_{d}$
and let  $g$ an element of $\widetilde G$ such that
$X' = gB$. $X$ and $X'$ are spacelike separated if   $$X_g= g^{-1} X= X_g(t, \psi,{\omega}) $$ belongs to the fundamental domain of $\widetilde {AdS}$  (global AdS itself, which is identified by  the condition $ |\Re t |< \pi$)
and  
\begin{equation}(X-X')^2 = (X_g-B)^2 < 2 - 2 \cosh\psi \cos t <0.
\end{equation}

\subsection{Quantum field theory}

When considering a scalar quantum field $\Phi(X)$  on the (covering of the) AdS manifold 
we will always assume that it is AdS covariant and that it satisfies the  {\em local commutativity} property:
for $X$ and $X'$ space-like separated the fields $\Phi(X)$ and $\Phi(X')$ commute.

The second property that we will assume is about the spectrum of the energy operator. 
The simplest way to state this condition is to identify the energy with the generator $M_{0d}$ of the rotations in the $(0,d)-$plane and demand that   
$M_{0d}$ be represented in the Hilbert space of the theory by a self-adjoint operator $\widehat M_{0d}$
bounded from below (for a more covariant statement see \cite{bem}).
The spectrum condition implies that there are two distinguished complex domains \cite{bem} of 
$CAdS_{d}$, invariant under real AdS transformations, where the 
two-point functions are holomorphic:
\begin{equation}
T^\pm =\{ Z    \in CAdS_{d},  \ \  \Im Z^2>0,\ \  \epsilon(Z) =  \mbox{sign}  (Y^0 X^{d}- X^0 Y^{d}) = \pm\}.
\label{Tubes}
\end{equation}
The following \underline{\em normal analyticity property} is equivalent to the positivity of the spectrum of the energy operator:
\begin{center}{\em  $W\left( X, X'\right)$
is the boundary value of a  function $ W\left(Z, Z'\right) $ holomorphic in  $\widetilde{T}^-\times \widetilde{T}^+$} \end{center}
where $\widetilde{T}^\pm $ are the coverings of the above chiral tubular domains.
By using  the complex  coordinates (see above)  the domains (\ref{Tubes}) are seen to be semi-tubes invariant under translations in the variable $\Re t$. They can be  described by the following inequalities:
 \begin{equation}\pm \sinh \Im t>  \left[\frac{(\sin\Im \psi)^2 + \left( (\cosh\Re  \psi)^2 - (\cos \Im \psi)^2 \right)(\Im \omega)^2}{  (\cosh \Re \psi)^2 - (\sin\Im \psi)^2}\right]^\frac{1}{2}.
 \label{retubes}\end{equation}
To grasp more intuitively the meaning of the above statement, in the case of real $\psi$ and $\omega$ i.e. when only the time coordinate $t$ is complexified it simply amounts 
to require that $W\left[Z(t, \psi,{\omega}), \, Z'(t', \psi',{\omega'})\right]$ has an analytic continuation to complex pairs such that  $\Im t<0$ and $\Im t'>0$ (flat tubes); this analyticity property is equivalent to the positivity of the energy spectrum.

AdS invariance further implies that
$W(Z,Z')$ is actually a function $w(\zeta)$
of a single complex variable $\zeta$ which  can be identified with
$Z\cdot Z'$ when  $Z$ and $Z'$ are both in the fundamental  sheet of $\widetilde{CAdS}_{\dd}$;
therefore AdS invariance and the spectrum condition together imply the following 
\underline{\em maximal analyticity property}:
\begin{center}
{\em $w(\zeta)$
is analytic on the covering $\widetilde\Theta$ of the
cut-plane $\Theta = \{{\mathbb C} \setminus[-1,1]\}$.} 
\end{center}
  For fields periodic in the time coordinate (semi-integer $\nu$ case) $w(\zeta)$ is in fact analytic in $\Theta$ itself.

One can now introduce all the standard Green functions.
The permuted two-point  function ${W}(X',X)$
is the boundary value of
${ W}(Z,Z')$ from the opposite domain
$\{(Z,Z'): Z\in { \widetilde  T}^{+}, \; Z'\in {\widetilde  T}^{-}\}$.
The retarded propagator  ${R}(X,X')$ is introduced by
splitting the support
of the commutator ${C}(X,X')$ as usual ($X,X' \in AdS_\dd$ real)
\begin{eqnarray}
&& C(X,X')={W}(X,X')-{W}(X',X), \\ 
&& {R}(X,X')= i\theta (t - t') { C}(X,X').
\end{eqnarray}
Finally, the chronological (Feynman) propagator is  given by
\begin{equation}
{\tilde{F}} (X,X')= -i \theta (t - t') {W}(X,X')- i \theta (t' - t) {W}(X',X).
\label{chronological}
\end{equation}
This definition refers to the global time-ordering of the 
covering  manifold. The uncovered AdS manifold is not globally time ordered. We take into account this property by modifying the previous definition as follows 
\begin{equation}
{F} (X,X')= -i \theta\left[\sin(t - t')\right]\, {W}(X,X')- i 
\theta\left[\sin(t' - t)\right]\, {W}(X',X).
\label{chronologicalads}
\end{equation}
The two prescriptions coincide on the fundamental sheet $|t-t'|<\pi$.

\subsection{Klein--Gordon fields}

In  $d$-spacetime dimensions the Wightman function having the above maximal analitycity property may be written in terms of  the associated Legendre's function of the second-kind:
\begin{equation}
W_{\nu}(Z,Z') = w_\nu(\zeta) =
\frac {e^{-i\pi\frac {d-2}2}}{(2\pi)^{\frac{d}2}}
(\zeta^2-1)^{-\frac {d-2}4} Q^{\frac {d-2}2}_{\nu-\frac 1 2}(\zeta).
\label{kgtp}
\end{equation}
In terms of the hypergeometric function
{\begin{eqnarray}  
w_\nu(\zeta) & = &\frac{\Gamma\left(\frac {d-1}2+\nu\right)\Gamma\left(\frac {d-1} 2 -\nu\right)}{ 2\,  {(2\pi)^{\frac{d}2}}\Gamma(\frac {d}2)}
\left[  \left(\frac{1}{\zeta-1}\right)^{\frac {d-2}2} \phantom{|}_1F_2\left(\frac 1 2-\nu ,\frac 1 2+\nu,\frac {d}2;\frac{ 1+ \zeta}2\right)\right. \cr&& - \left. 
e^{\mp i(\nu-\frac 1 2)\pi} 
\left(\frac{1}{\zeta+1}\right)^{\frac {d-2}2}  \phantom{|}_1F_2\left(\frac 1 2-\nu,\frac 1 2+\nu,\frac {d}2;\frac{ 1- \zeta}2\right)\right],\label{kgtp3}
\end{eqnarray}
where the upper or lower sign in the phase is chosen according as $\Im \zeta>  0$ or $\Im \zeta<  0$} \cite{Bateman}. 
Here the parameter $\nu$  is related to the mass as follows
\begin{equation}
\nu^2  = \frac {(d-1)^2} {4} + \left(m \, R\right)^2,
\label{nu}
\end{equation}
and the normalization of $W_\nu$ is chosen by imposing
the short-distance Hadamard behavior.

{For $d=4$ the coefficient at the RHS of Eq. (\ref{kgtp3}) is divergent when  $\nu =  \frac 32 , \frac 52,\ldots  $}. At the same time for $\nu =  \frac 32 , \frac 52,\ldots $, the phases multiplying the  second term are equal in the upper and lower half-planes. The hypergeometric series become sums and  the constant terms exactly cancel. Therefore for  $\nu = \frac 12,  \frac 32 , \frac 52 , \ldots  $, etc. we may 
extract two-point functions which are  periodic in the time variable (and therefore  live on the true AdS manifold) and  analytic in the cut plane $\Theta$. 
This may be done by analytic continuation in the dimension $d$. For instance  for $\nu = \frac 32$ (the massless $m=0$) case we have
\begin{eqnarray}  
 w^d_{\frac 32}(\zeta) = \frac{\Gamma\left(\frac {d-1}2-\frac 32 \right) \Gamma\left(\frac {d-1} 2 +\frac 32\right)}{ 2\,{(2\pi)^{\frac{d}2}} \Gamma(\frac {d}2)}
\left[  \left(\frac{1}{\zeta  -1 }\right)^{\frac {d-2}2}
  \left(1- \frac{4}{d}\frac {1+\zeta}2 \right)\right. \cr + \left. 
\left(\frac{1}{\zeta+1}\right)^{\frac {d-2}2}  \left(1 - \frac  4 {d} \frac {1-\zeta }2  \right)\right] .
\end{eqnarray}
Taking the limit at  $d=4$ we get
\begin{eqnarray}  
 w_{\frac 3 2}(\zeta) 
=-\frac{1}{ 2{(2\pi)^{2}} }
 \log \left(\frac {\zeta+1}{\zeta  -1}\right)+
\frac{ 1}{ 2\,(2\pi)^{2}}
\left[  \left(\frac{1}{\zeta -1 }\right)
 +
\left(\frac{1}{\zeta+1}\right) \right]. 
\end{eqnarray}
Note that for  $\nu = \frac 12$ the coefficient at the RHS of Eq. (\ref{kgtp3})  is finite; both hypergeometric series are equal to 1 and the RHS is just the difference of the two poles:
\begin{eqnarray}  
 w_{\frac 1 2}(\zeta) 
=\frac{ 1}{ 2\,(2\pi)^{2}}
\left[  \left(\frac{1}{\zeta -1 }\right)
 -
\left(\frac{1}{\zeta+1}\right) \right] . \label{12}
\end{eqnarray}

\subsection{Chronological propagators}

Let us first compute the chronological propagator by the above prescriptions when the first point $X$ belong to the fundamental domain of the $\widetilde {AdS}$ space with $-\pi< t<\pi$, and we take the second point at the origin $X' = B$ (so that $t'=0$).

In accordance with both Eqs. (\ref{chronological}) and (\ref{chronologicalads}) and the normal analiticity property, when $0< t<\pi$ the real event $X(t, \psi,{\omega})$ has to be taken at the boundary of $T^-$ so that
$$\zeta = \cosh \psi\cos (t-i \epsilon ) = \xi+ i \epsilon \sin t, \ \ \ \ \epsilon>0.$$ 
On the other hand when $-\pi < t<0$  the real event $X(t, r,{\omega})$  has to be considered at the boundary of $T^+$; 
$$\zeta = \cosh\psi\cos (t+i \epsilon ) =\xi - i \epsilon \sin t, \ \ \ \ \epsilon>0.$$
Therefore  as long as we have $-\pi <t<\pi$ we always have $\zeta = X\cdot B + i \epsilon|\sin t| $ with $\epsilon>0$ and 
\begin{equation}
{F} (X,B) = {-i w}\left( \xi+ i \epsilon |\sin t|\right) \ \ \ -\pi<t< \pi.
\label{chronological2}
\end{equation}
Hence, the  equation
\begin{equation}
{F} (X,X') = {-i w}( \xi+ i \epsilon) \ 
\label{chronological2bis}
\end{equation}
provides the time-ordered propagator whenever $X,X'\in \widetilde{AdS}$ 
are such that there exists a  $g$ inf $\widetilde G$ such that
$X' = gB$  and $g^{-1}X'$ is in the fundamental domain of $\widetilde {AdS}$  (identified by  the condition $ |\Re t |< \pi$). 
It requires the spectrum condition to hold. 
By applying Eq. (\ref{chronological2bis}) to the Wightman function (\ref{kgtp3}) we get the Feynman propagator  form presented in Eq. (\ref{FAdS}).
In the simplest case  $\nu = \frac 12$  Eq. (\ref{chronological2bis}) gives 
\begin{eqnarray}  
  F_{\frac 1 2}(X,X') = f_{\frac 1 2}(\xi) 
=-\frac{ i}{ 2\,(2\pi)^{2}}
\left[  \left(\frac{1}{\xi -1 +i \epsilon }\right)
 -
\left(\frac{1}{\xi+1  +i \epsilon}\right) \right] .
\end{eqnarray}
What is the situation in $\widetilde{AdS}$? One possibility is to insist in using the same $i \eps$ prescription as above. The other possibility seems obvoius: try to use the global ordering of the covering manifold. This choice however  forces to move to the covering manifold $\widetilde {AdS}$ even in the periodic cases. 
Let us illustrate this phenomenon 
by applying  the global definition (\ref{chronological}) to the case $\nu = \frac 12$; we get the following propagator:

\begin{equation} F_{\frac 1 2}(X,X') = \frac{ i}{ 2\,(2\pi)^{2}} \left\{ \begin{array} {lcl}

  \left(\frac{1}{\xi -1 + i \epsilon}\right)
 -
 \left(\frac{1}{\xi +1+i\epsilon }\right)    & \makebox{ for } & n<\frac{|t-t'|}{2\pi}<n+\frac 1 2   \cr 

  \left(\frac{1}{\xi -1 - i \epsilon}\right)
 -
 \left(\frac{1}{\xi +1-i\epsilon }\right)   & \makebox{ for } &n+\frac 1 2<\frac{|t-t'|}{2\pi} <n+1 
\end{array}\right. n= 0 ,1, 2,\ldots \end{equation} This relation remains true for all periodic cases Klein-Gordon fields with half-integer $\nu= k + \frac {1}2$: 
\begin{equation} F_{\frac {k+1} 2}(X,X') =  \left\{ \begin{array} {lcl}
-iw(\xi+i\epsilon) & \makebox{ for } & n<\frac{|t-t'|}{2\pi}<n+\frac 1 2   \cr 
-iw(\xi-i\epsilon)   & \makebox{ for } &n+\frac 1 2<\frac{|t-t'|}{2\pi} <n+1
\end{array}\right. n= 0 ,1, 2,\ldots \end{equation}
Correctly understood, this last formula is indeed valid also in the non periodic case. In particular in $d=4$ for Klein-Gordon fields this gives 
\begin{equation} 
F_{ {\nu} }(X,X') = 
 \left\{ \begin{array} {lcl}
 \frac {i e^{2 i \pi n(\nu -\frac 12)\makebox{\tiny{sgn}}(t-t')}}{(2\pi)^{2}} \bigl[(\xi+i\epsilon)^2-1\bigr]^{-\frac {1} 2} Q^{1}_{\nu-\frac 1 2}(\xi+i\epsilon)
& \makebox{ for } & n<\frac{|t-t'|}{2\pi}<n+\frac 1 2 \cr 
\frac {ie^{2 i \pi n (\nu -\frac 12) \makebox{\tiny{sgn}}(t-t')}}{(2\pi)^{2}}  \bigl[(\xi-i\epsilon)^2-1\bigr]^{-\frac {1} 2} Q^{1}_{\nu-\frac 1 2}(\xi-i\epsilon)
& \makebox{ for } &n+\frac 1 2<\frac{|t-t'|}{2\pi} <n+1 \cr
\end{array}\right. \label{cpcov}\end{equation}
Here  $\xi \in \Theta$ is the scalar product  of the projections of the two points $X,X'\in \widetilde{AdS}$ on the  fundamental sheet. The obtained $i \eps$ prescription in $\widetilde{AdS}$ is Feynman's at coinciding points and is Wightman's in elsewhere and this gives the alternating signs of $i\eps$ as one crosses the cuts.

\bibliographystyle{ssg}
\bibliography{ads}

\end{document}